\documentclass[conference]{IEEEtran}
\usepackage{amsthm,amsmath,amssymb,mathtools,bm,etoolbox,float,hyperref}
\usepackage[dvipsnames]{xcolor}
\usepackage{graphicx,cite,cuted}
\usepackage{bigints}
\usepackage{caption}

\usepackage{stackengine}

\usepackage{balance}

\theoremstyle{corollary}
\newtheorem{corollary}{Corollary}

\usepackage{mathrsfs,verbatim}
\usepackage{ellipsis}
\usepackage{tikz}
\usepackage{tikz-3dplot}
\usepackage[utf8]{inputenc}
\usepackage{pgfplots} 
\usepackage{pgfgantt}
\usepackage{pdflscape}
\pgfplotsset{compat=newest} 
\pgfplotsset{plot coordinates/math parser=false}
\pgfplotsset{every  tick/.style={black,},ylabel style={font=\tiny},xlabel style={font=\tiny},tick label style={font=\tiny},legend style= {font=\scriptsize},
minor x tick num=1,minor y tick num=1,xminorticks=true,yminorticks=true,}
  \newlength\fheight
\newlength\fwidth

\usepackage{xinttools} % for the \xintFor***
\usepackage{tikz}
\usepackage{tkz-euclide}
\usetikzlibrary{calc}
\newtheorem{theorem}{Theorem}
\newtheorem{proposition}{Proposition}
\newtheorem{lemma}{Lemma}
\newtheorem{remark}{Remark}

\def\biglen{20cm} % playing role of infinity (should be < .25\maxdimen)
\tikzset{
  half plane/.style={ to path={
       ($(\tikztostart)!.5!(\tikztotarget)!#1!(\tikztotarget)!\biglen!90:(\tikztotarget)$)
    -- ($(\tikztostart)!.5!(\tikztotarget)!#1!(\tikztotarget)!\biglen!-90:(\tikztotarget)$)
    -- ([turn]0,2*\biglen) -- ([turn]0,2*\biglen) -- cycle}},
  half plane/.default={1pt}
}

\DeclareMathAlphabet{\pazocal}{OMS}{zplm}{m}{n}

\usepackage{caption}
\usepackage{multicol,subcaption}
\IEEEoverridecommandlockouts
\usepackage{cite}
\usepackage{amsmath,amssymb,amsfonts}
\usepackage{algorithmic}
\usepackage{graphicx}
\usepackage{textcomp}
\usepackage{xcolor}
\def\BibTeX{{\rm B\kern-.05em{\sc i\kern-.025em b}\kern-.08em
    T\kern-.1667em\lower.7ex\hbox{E}\kern-.125emX}}
\IEEEoverridecommandlockouts\IEEEpubid{\makebox[\columnwidth]{ 978-1-6654-3540-6/22~\copyright~2022 IEEE \hfill} \hspace{\columnsep}\makebox[\columnwidth]{ }}

\begin{document}

\title{Full-Duplex Massive MIMO Cellular Networks with Low Resolution ADC/DAC}

\author{\IEEEauthorblockN{Elyes Balti and Brian L. Evans}\thanks{This work was supported by NVIDIA, an affiliate of the 6G@UT center within the Wireless Networking and Communications Group at The University of Texas at Austin.}
\IEEEauthorblockA{6G@UT Research Center, Wireless Networking and Communications Group (WNCG) \\ The University of Texas at Austin, Austin, TX\\
ebalti@utexas.edu, bevans@ece.utexas.edu}
}

\maketitle

\begin{abstract}
In this paper, we provide an analytical framework for full-duplex (FD) massive multiple-input multiple-output (MIMO) cellular networks with low resolution analog-to-digital and digital-to-analog converters (ADCs and DACs). Matched filters are employed at the FD base stations (BSs) at the transmit and receive sides. For both reverse and forward links, our contributions are (1) derivations of the signal-to-quantization-plus-interference-and-noise ratio (SQINR) for general and special cases;
(2) derivations of spectral efficiency for asymptotic cases as well as for power scaling laws; and
(3) quantifying effects of quantization error, loopback self-interference, and inter-user interference for hexagonal cells and Poisson Point Process (PPP) tessellations on outage probability and spectral efficiency.
\end{abstract}


\section{Introduction}
Full-duplex (FD) has emerged as an attractive solution to double the spectral efficiency because transmission and reception occur simultaneously in the same resource blocks. In addition, FD devices can use a shared array; i.e., the same array can be used for transmission and reception instead of having two separate arrays, which can substantially reduce cost. These benefits make FD attractive in practice, e.g. machine-to-machine communications and integrated access and backhaul proposed in 3GPP release 17 \cite{release17}. Although FD brings many advantages, it suffers from loopback self-interference (SI) caused by the simultaneous transmission and reception in the same resource blocks. This loopback signal cannot be neglected as the relative SI power can be several orders of magnitude stronger than the received signal power, which can render FD systems dysfunctional \cite{ianmagazine,zf}.  

\textcolor{black}{Massive MIMO has been proposed as a viable solution to enable FD operation. A massive number of antennas would provide significant multiplexing and diversity gains and serve a larger number of users simultaneously \cite{f2}.  In a rich scattering environment, a high data rate can be achieved without increasing bandwidth by utilizing the Degrees of Freedom (DoF) that come with a massive number of antennas \cite{ianjournal} in addition to the time and frequency dimensions.  A large number of antennas can also minimize SI as well as intra and inter-cell interference by narrowing and focusing radiated energy towards the intended user equipment (UE) location.}

\textcolor{black}{In massive MIMO FD systems, one of the most challenging problems is interference cancellation. Interference cancellation is highly dependent on the accuracy of channel state information (CSI) at the base stations (BSs). To tackle this problem, many related papers consider massive MIMO in Time Domain Duplexing (TDD) mode because of channel reciprocity instead of using Frequency Domain Duplexing (FDD). Full-duplex massive MIMO has been considered for cellular networks, millimeter wave (mmWave) IEEE 802.11ad and 802.11ay Wi-Fi standards, and 5G New Radio in 3GPP Release 15 \cite{overview,massiveforward}.}

Employing a massive number of antennas, however, can lead to \textcolor{black}{high} power consumption, esp. for full-digital systems wherein each RF chain and analog-to-digital and digital-to-analog converters (ADCs and DACs) \textcolor{black}{is} dedicated to an antenna element. In addition, power consumption can be prohibitive for the \textcolor{black}{larger} bandwidths in \textcolor{black}{mmWave} applications. For this reason, low-resolution ADCs/DACs have been proposed to reduce power consumption at the expense of spectral efficiency \cite{massiveadc,R3,fdvehicular}. Increasingly, energy efficiency is becoming a more important system design measure than spectral efficiency. 

In this context, we consider FD with low-resolution ADCs and DACs in cellular networks. The BS operates in FD mode while UE operates in half-duplex (HD) mode. Due to the limited space, we defer analysis of pilot contamination to future work, and for now, assume pilot orthogonality is maintained across the network cells. Communication performance is simulated for a hexagonal lattice with different tiers and a Voronoi tessellation for PPP networks. To the best of our knowledge, this is the first work that considers FD systems with low resolution data converters for cellular networks.

Section II discusses the network model. Sections III and IV analyze reverse and forward links. Asymptotic analysis and power scaling laws are given in Section V, and numerical results appear in Section VI. Section VII concludes the paper.

{\bf Notation}. Italic non-bold letters refer to scalars while bold lower and upper case stand for vector and matrix, respectively. We denote subscripts $u$ for uplink and $d$ for downlink.

\section{Network Model}
\begin{figure}
    \centering
    \includegraphics[scale=0.7]{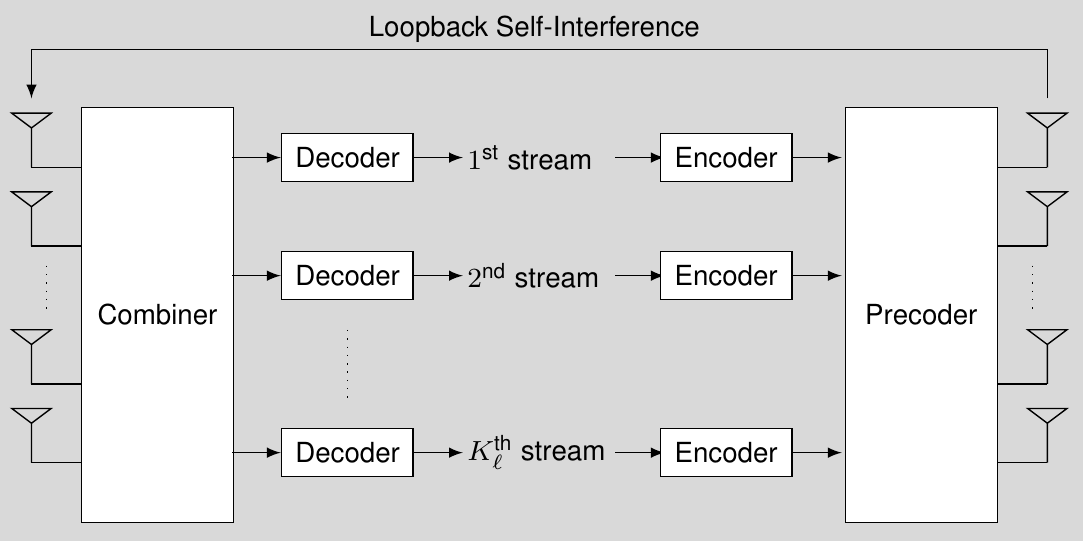}
    \caption{Full-duplex BS. Uplink UE sends data to the BS independently from the data intended for the downlink UE sent from the BS. Since the BS transmits and receives at the same time and on the same resource blocks, SI leakage creates loopback interference from the TX to the RX side of the BS.}
    \label{system}
\end{figure}
We consider a macrocellular network where each BS operates in FD mode and is equipped with $N_{\mathsf{a}} \gg 1$ antennas. Each UE operates in HD mode and has a single antenna.

\subsection{Large-Scale Fading}
Each UE is associated with the BS from which it has the highest large-scale channel gain. We denote $K_\ell^u$ and $K_\ell^d$ to be the number of uplink and downlink UEs served by the $\ell$-th BS. The large-scale gain between the $\ell$-th BS and the $k$-th user connected to the $l$-th BS comprises pathloss with exponent $\eta > 2$ and independently and identically distributed (IID) shadowing across the paths, and is defined as
\begin{equation}
G_{\ell,(l,k)} = \frac{L_{\text{ref}}}{r_{\ell,(l,k)}^\eta} \chi_{\ell,(l,k)}  
\end{equation}
where $L_{\text{ref}}$ is the pathloss intercept at a unit distance, $r_{\ell,(l,k)}$ the link distance, and $\chi_{\ell,(l,k)}$ as the shadowing coefficient satisfying $\mathbb{E}[\chi^\delta] < \infty$, where $\delta = 2/\eta$. 

Without loss of generality, we denote the 0-th BS as the focus of interest and we drop its subscript.

Further, we introduce large-scale fading between the UEs to model the inter-user communication. We denote $T_{(\ell,n),(l,k)}$ as the large-scale channel gain between the $n$-th user and the $k$-th user associated with the $\ell$-th and $l$-th BSs, respectively.

\subsection{Small-Scale Fading}
We denote $\boldsymbol{h}_{\ell,(l,k)} \sim \mathcal{N}_{\mathbb{C}}(\bold{0},\boldsymbol{I})$ as the normalized reverse link $N_{\mathsf{a}} \times 1$ small-scale fading between the $k$-th user located in cell $l$ and the BS in cell $\ell$ and $\boldsymbol{h}^*_{\ell,(l,k)}$ as the forward link reciprocal, assuming time division duplexing (TDD) with perfect calibration \cite{massiveforward}. In addition, we denote $\boldsymbol{g}_{(\ell,n),(l,k)} \sim \mathcal{N}_{\mathbb{C}}(0,\sigma_{\mathsf{iui}}^2)$ as $1 \times 1$ small-scale fading between the $n$-th user in cell $\ell$ and the $k$-th user in cell $l$. We further denote $\boldsymbol{H}_{\mathsf{SI}} \sim \mathcal{N}_{\mathbb{C}}(\bold{0},\boldsymbol{\mu}_{\mathsf{SI}}^2)$ as the SI channel matrix ($N_{\mathsf{a}} \times N_{\mathsf{a}}$) \cite{massiveadc}.

\begin{remark}
We assume perfect SI channel knowledge since the SI channel has a line-of-sight (LOS) component (near-field) that is dominant and deterministic and an external non-LOS component that is negligible and random. Without loss of generality, we consider a random SI channel matrix that can be estimated by Least Squares or other conventional methods. If the channel matrix is sparse, a compressive sensing method will be the best candidate, esp. for a large matrix dimension. In addition, we denote the SI channel estimate as $\hat{\boldsymbol{H}}_{\mathsf{SI}}$ and include the estimation error as an additional source of SI. SI channel estimation is out of the scope of this work.
\end{remark}

\subsection{Quantized Signal Model}
For infinite resolution, a typical received signal $\boldsymbol{y}$ is
\begin{equation}\label{unqsignal}
    \boldsymbol{y} = \boldsymbol{H}\boldsymbol{x} + \boldsymbol{n}
\end{equation}
where $\boldsymbol{H}$, $\boldsymbol{x}$, and $\boldsymbol{n}$ are the channel matrix, precoded symbols and additive white Gaussian noise (AWGN), respectively. Several nonlinear quantization models have been proposed; however, their analysis is complex for a higher number of ADC bits. In quantized systems, a lower bound for spectral efficiency has been derived by treating quantization as additive Gaussian noise with variance inversely proportional to the quantizer resolution, i.e. $2^{-b}$ times the received input power where $b$ is the number of ADC bits. An additive quantization noise model (AQNM) has recently been proposed for mmWave signals with an arbitrary number of ADC bits \cite{R5,R6,R7}. In addition, \cite{lowres} derived Gaussian approximations using Bussgang Theory to linearize the nonlinear quantization distortion which is similar to the AQNM. The received signal (\ref{unqsignal}) is processed through the RF chains and then converted to the digital discrete-time domain by the ADC. The AQNM represents the quantized version of (\ref{unqsignal}) given by
\begin{equation}\label{quansignal}
\boldsymbol{y}_q = \alpha \boldsymbol{y} + \boldsymbol{q}
\end{equation}
where $\boldsymbol{q}$ is the additive quantization noise, $\alpha = 1-\rho$ and $\rho$ is the inverse of the signal-to-quantization-plus-noise ratio (SQNR), which is inversely proportional to the square of the resolution of an ADC, i.e., $\rho = \frac{\pi\sqrt{3}}{2} \cdot 2^{-2b}$:

\vspace*{0.1in}
\begin{tabular}{cccccc}
$\boldsymbol{b}$ & 1 & 2 & 3 & 4 & 5\\
\hline
$\boldsymbol{\rho}$ & 0.3634 & 0.1175 & 0.03454 & 0.009497 & 0.002499
\end{tabular}
 
%\begin{table}[tbhp]
%\renewcommand{\arraystretch}{1}
%\caption{$\rho$ for different values of $b~(b \leq 5)$ \cite{22}.}
%\label{etaparam}
%\centering
%\begin{tabular}{cccccc}
%$\boldsymbol{b}$ & 1 & 2 & 3 & 4 & 5\\
%\hline
%$\boldsymbol{\rho}$ & 0.3634 & 0.1175 & 0.03454 & 0.009497 & 0.002499
%\end{tabular}
%\end{table}

\textcolor{black}{\subsection{Full-Duplex and Low Resolution ADC/DAC}
Without loss of generality, we analyze the FD case for low resolution ADC/DAC in single-cell single-user scenario. Then we generalize the analysis for cellular network for reverse and forward links.
The received signal $\boldsymbol{y}^u$ at the BS is given by
\begin{equation}
\boldsymbol{y}^u = \sqrt{P_u}\boldsymbol{H}_u\boldsymbol{s}_u + \sqrt{P_d}\boldsymbol{H}_{\mathsf{SI}}\boldsymbol{x}_d + \boldsymbol{n}_u  
\end{equation}
with the downlink unquantized precoded signal $\boldsymbol{x}_d$ given by
\begin{equation}
\boldsymbol{x}_d = \boldsymbol{F}\boldsymbol{s}_d     
\end{equation}
where $\boldsymbol{F}$ is the precoder. As the AQNM approximates the quantization error, the received $\boldsymbol{y}^u_q$ and transmitted signals $\boldsymbol{x}_{d,q}$ after the ADC/DAC at the BS can be obtained by
\begin{equation}
\boldsymbol{y}^u_q = \alpha_u \boldsymbol{y}^u + \boldsymbol{q}_u    
\end{equation}
\begin{equation}
   \boldsymbol{x}_{d,q} = \alpha_d \boldsymbol{x}_d + \boldsymbol{q}_d    
\end{equation}
We define the AQNM covariance matrices as follows \cite{massiveadc}
\begin{equation}
\boldsymbol{R}_{\boldsymbol{q}_u} = \mathbb{E}[\boldsymbol{q}_u\boldsymbol{q}_u^*] = \alpha_u(1-\alpha_u)\text{diag}\left(P_u\boldsymbol{H}_u\boldsymbol{H}_u^* + \boldsymbol{Q} + \sigma^2 \boldsymbol{I}_{N_{\mathsf{a}}}  \right)    
\end{equation}
\begin{equation}
\boldsymbol{R}_{\boldsymbol{q}_d} = \mathbb{E}[\boldsymbol{q}_d\boldsymbol{q}_d^*] = \alpha_d(1-\alpha_d)\text{diag}\left(\boldsymbol{F}\boldsymbol{F}^* \right)
\end{equation}
where $\boldsymbol{Q}$ is given by
\begin{equation}
\boldsymbol{Q} = P_u\boldsymbol{H}_{\mathsf{SI}}\left( \alpha_d^2\boldsymbol{F}\boldsymbol{F}^* +\boldsymbol{R}_{\boldsymbol{q}_d}  \right)    \boldsymbol{H}_{\mathsf{SI}}^*
\end{equation}}

\section{Reverse Link (Uplink) Analysis}
The $k$-th uplink user in cell $\ell$ sends data symbol $s_{\ell,k} \sim \mathcal{N}_{\mathbb{C}}(0,1)$. Since the BS operates in FD mode, the signal transmitted to downlink users leaks to the BS receive array; the loopback SI corrupts the uplink user signals. The BS of interest observes the following quantized received signal vector
\begin{equation}
\label{uplink1}
\begin{split}
\boldsymbol{y}_q^u =& \alpha_u\sum_\ell \sum_{k=0}^{K^u_\ell-1} \sqrt{G_{\ell,k}P_{\ell,k}}  \boldsymbol{h}_{\ell,k} s_{\ell,k} + \alpha_u\sqrt{P_{\mathsf{SI}}} \boldsymbol{H}_{\mathsf{SI}}\boldsymbol{q}_d\\&+ \alpha_u\alpha_d\sqrt{P_{\mathsf{SI}}}\sum_{k=0}^{K^d-1} \boldsymbol{H}_{\mathsf{SI}}\boldsymbol{f}_{k} s_{k}^d  + \boldsymbol{q}_u + \alpha_u\boldsymbol{v}     
\end{split}
\end{equation}
where $P_{\ell,k}$ is the transmit power of uplink user $k$ served by the $\ell$-th BS and $\boldsymbol{v} \sim \mathcal{N}_{\mathbb{C}}(0,\boldsymbol{I})$ is the $N_{\mathsf{a}} \times 1$ AWGN vector.

\subsection{Channel Hardening}
One of the benefits of having a massive number of antennas is the hardening of the filtered signals. Suppose that, rather than $\boldsymbol{w}^*_{k} \hat{\boldsymbol{h}}_{k}$ (where $\hat{\boldsymbol{h}}$ is the estimate of $\boldsymbol{h}$), the decoder regards $\mathbb{E}[\boldsymbol{w}^*_k\boldsymbol{h}_{k}]$ as the filtered channel. The receiver can compute this value from the channel statistics while the fluctuation of the filtered signal around the mean can be treated as self-interference. By decomposing the interference into inter-cell and intra-cell and applying the linear receive filter $\boldsymbol{w}^*_k$ at the $k$-th user ($y_k = \boldsymbol{w}^*_k\boldsymbol{y}$), the received signal at the BS of interest of the $k$-th user is given by (\ref{uplink2}) on the next page.

\begin{figure*}[t]
\begin{equation}\label{uplink2}
\begin{split}
y^u_{q,k} =& \underbrace{\alpha_u\sqrt{G_{k}P_k} \mathbb{E}[\boldsymbol{w}^*_k\boldsymbol{h}_{k}]s_k}_{\textsf{Desired Signal}} + \underbrace{\alpha_u \sum_{\text{k}\neq k} \sqrt{G_{\text{k}}P_{\text{k}}}\boldsymbol{w}^*_{k} \boldsymbol{h}_{\text{k}}s_{\text{k}}}_{\textsf{Intra-Cell Interference}}+\underbrace{\alpha_u\sum_{\ell\neq 0}\sum_{\text{k}=0}^{K^u_\ell-1}\sqrt{G_{\ell,\text{k}}P_{\ell,\text{k}}}\boldsymbol{w}^*_{k}\boldsymbol{h}_{\ell,\text{k}}s_{\ell,\text{k}}}_{\textsf{Inter-Cell Interference}}+ \underbrace{\alpha_u \boldsymbol{w}^*_k\boldsymbol{v}}_{\textsf{Filtered Noise}}
\\&+\overbrace{ \underbrace{\alpha_u \sqrt{G_{k}P_k}\left( \boldsymbol{w}^*_k\boldsymbol{h}_{k} - \mathbb{E}[\boldsymbol{w}^*_k\boldsymbol{h}_{k}] \right) s_k}_{\textsf{Channel Estimation Error}}   + \underbrace{\alpha_u\alpha_d\sqrt{P_{\mathsf{SI}}}\sum_{\text{k}=0}^{K^d-1}\boldsymbol{w}^*_k\boldsymbol{H}_{\mathsf{SI}}\boldsymbol{f_{\text{k}}}s^d_{\text{k}}}_{\textsf{Self-Interference due to Full-Duplexing}}}^{\textsf{Aggregate Self-Interference}} + \underbrace{\alpha_u\sqrt{P_{\mathsf{SI}}}\boldsymbol{w}^*_k\boldsymbol{H}_{\mathsf{SI}}\boldsymbol{q}_d + \boldsymbol{w}^*_k\boldsymbol{q}_u}_{\textsf{Aggregate AQNM}}
\end{split}
\end{equation}
\vspace*{-.5cm}
\end{figure*}

\subsection{Matched Filter Receiver}
We adopt the matched filter receiver to design the combiner $\boldsymbol{w}_k,~k=0,\ldots,K^u-1$. 
%A matched filter for user $k$ satisfies $\boldsymbol{w}^{\text{MF}}_k \propto \hat{\boldsymbol{h}}_k $. 
%The scaling is important to operate the decoder, but immaterial since it equally affects the received signal as well as the noise the interference. By neglecting the pilot contamination, the linear receive matched filter of the $k$-th user at the base station of interest is given by (\ref{uplink2})
%\begin{equation}
%\boldsymbol{w}^{\text{MF}}_k = \sqrt{\frac{\frac{P_k}{P_u}\textsf{SNR}_k^u}{ 1 + \frac{P_k}{P_u}\textsf{SNR}_k^u    }} \boldsymbol{h}_k   
%\end{equation}
\begin{corollary}\label{cor1}
The matched filter receiver $\boldsymbol{w}^{\mathsf{MF}}_k$ (transmitter $\boldsymbol{f}^{\mathsf{MF}}_k$) has the following properties
\begin{enumerate}
    \item $\mathbb{E}\left[\|\boldsymbol{w}^{\mathsf{MF}}_k \|^2 \right] =\mathbb{E}\left[\|\boldsymbol{f}^{\mathsf{MF}}_k \|^2 \right] = N_{\mathsf{a}}$.
    \item $\mathbb{E}\left[\| \boldsymbol{w}^{\mathsf{MF}}_k \|^4 \right] = \mathbb{E}\left[\| \boldsymbol{f}^{\mathsf{MF}}_k \|^4 \right]= N_{\mathsf{a}}^2 + N_{\mathsf{a}}$.
    \item $\mathbb{E}\left[ \left| \boldsymbol{w}^{\mathsf{MF}*}_k \boldsymbol{h}_{\ell,\mathrm{k}} \right|^2 \right] = \mathbb{E}\left[ \left|\boldsymbol{h}^*_{\ell,k} \boldsymbol{f}^{\mathsf{MF}}_{\ell,\mathrm{k}}  \right|^2 \right] = N_{\mathsf{a}}$.
    %\item $\mathbb{E}\left[ \boldsymbol{w}^{\mathrm{MF}*}_k \boldsymbol{h}_{\text{k}} \right]  = \left| \mathbb{E}\left[ \boldsymbol{h}^*_{\text{k}} \boldsymbol{f}^{\mathrm{MF}}_k  \right]   \right|^2 = N_{\mathsf{a}}^2$.
\end{enumerate}
\end{corollary}
\begin{theorem}
The output SQINR of the $k$-th uplink user is (\ref{uplinksqinr}).
\begin{equation}\label{uplinksqinr}
\overline{\mathsf{sqinr}}_k^{\mathsf{MF}} = \frac{ \alpha_u^2 G_k P_k  \left|\mathbb{E}[\boldsymbol{w}^{\mathsf{MF}*}_k\boldsymbol{h}_k]\right|^2  }{\overline{\mathsf{den}}^{\mathsf{MF}}_u}    
\end{equation}
\end{theorem}
where $\overline{\mathsf{den}}^{\mathsf{MF}}_u$ is given by (\ref{den1}) on the next page.
\begin{figure*}[t]
\begin{equation}\label{den1}
\begin{split}
\overline{\mathsf{den}}^{\mathsf{MF}}_u =& \alpha_u^2G_kP_k\textsf{var}[\boldsymbol{w}_k^{\mathsf{MF}*}\boldsymbol{h}_k] + \alpha_u^2\sum_{\text{k}\neq k}G_{\text{k}}P_{\text{k}} \mathbb{E}\left[ \left| \boldsymbol{w}^{\mathsf{MF}*}_k \boldsymbol{h}_{\text{k}} \right|^2  \right] + \alpha_u^2 \sum_{\ell \neq 0}\sum_{\text{k}=0}^{K^d_{\ell}-1}G_{\ell,\text{k}}P_{\ell,\text{k}}\mathbb{E}\left[ \left| \boldsymbol{w}^{\mathsf{MF}*}_k \boldsymbol{h}_{\ell,\text{k}} \right|^2 \right]\\&+\alpha_u^2\sigma^2\mathbb{E}\left[ \left\| \boldsymbol{w}^{\mathsf{MF}}_k \right\|^2\right] + \alpha_u^2\alpha_d^2 P_{\mathsf{SI}}\sum_{\text{k}=0}^{K^d-1}\mathbb{E}\left[ \left| \boldsymbol{w}^{\mathsf{MF}*}_k \boldsymbol{H}_{\mathsf{SI}} \boldsymbol{f}_{\text{k}}^{\mathsf{MF}} \right|^2  \right] + \alpha_u^2 P_{\mathsf{SI}}\mathbb{E}\left[ \left| \boldsymbol{w}^{\mathsf{MF}*}_k\boldsymbol{H}_{\mathsf{SI}}\boldsymbol{q}_d \right|^2  \right] + \mathbb{E}\left[ \left| \boldsymbol{w}^{\mathsf{MF}*}_k\boldsymbol{q}_u \right|^2 \right]
\end{split}
\end{equation}
\vspace*{-.5cm}
\end{figure*}
\begin{equation}\label{a1}
\sum_{\text{k}=0}^{K^d-1}\mathbb{E}\left[ \left|\boldsymbol{w}_k^*\boldsymbol{H}_{\mathsf{SI}}\boldsymbol{f}_{\text{k}} \right|^2 \right] = \mu^2_{\mathsf{SI}}K^d N_{\mathsf{a}}^2    
\end{equation}
\begin{equation}\label{a3}
\begin{split}
 \mathbb{E}[ \left| \boldsymbol{w}_k^*\boldsymbol{q}_u\right|^2]  =&N_{\mathsf{a}} \alpha_u(1-\alpha_u)( 2G_kP_k + \sum_{\text{k}\neq k}G_{\text{k}}P_{\text{k}} \\& + \sum_{\ell \neq 0}\sum_{\text{k}}G_{\ell,\text{k}}P_{\ell,\text{k}} + \alpha_dP_{\mathsf{SI}}\mu_{\mathsf{SI}}^2N_{\mathsf{a}} + \sigma^2 )
\end{split}
\end{equation}
\begin{equation}\label{a2}
\mathbb{E}[ \left| \boldsymbol{w}_k^* \boldsymbol{H}_{\mathsf{SI}} \boldsymbol{q}_d \right|^2] = \alpha_d(1-\alpha_d)\mu^2_{\mathsf{SI}}N_{\mathsf{a}}^2
\end{equation}
The numerator of (\ref{uplinksqinr}) and the first four terms of (\ref{den1}) on the next page can be solved using the properties in Corollary \ref{cor1}, while the remaining terms of (\ref{den1}) can be derived as in (\ref{a1}), (\ref{a2}), and (\ref{a3}).
\begin{proof}
Proofs of (\ref{a1}), (\ref{a2}), and (\ref{a3}) are provided by Appendix A in \cite{massiveadc}.
\end{proof}
With $\overline{\mathsf{sqinr}}^{\mathsf{MF}}_k,~k=0,\ldots K^u-1$, stable over the respective local neighborhoods, the evaluation of the gross spectral efficiencies does not require averaging over the fading realizations, but rather is directly computed as
\begin{equation}
    \frac{\bar{\mathcal{I}}_k}{B} = \log\left( 1 + \overline{\mathsf{sqinr}}_k \right),~k=0,\ldots,K^u-1
\end{equation}
where $B$ is the bandwidth.
\begin{corollary}\label{cor2}
To further characterize the spectral efficiency, we can derive a new bound using the following formula. Assuming statistical independence between $x$ and $y$, we have
\begin{equation}
\mathbb{E}\left[ \log\left(1 + \frac{x}{y} \right) \right] \cong \log\left( 1 + \frac{\mathbb{E}[x]}{\mathbb{E}[y]} \right)    
\end{equation}
\end{corollary}
Assuming perfect channel state information (CSI), i.e., without channel hardening, and applying Corollary \ref{cor2}, the numerator of (\ref{uplinksqinr}) becomes
\begin{equation}
\overline{\mathsf{num}}^{\mathsf{MF}}_u = \alpha_u^2 G_k P_k  \mathbb{E}\left[\left|\boldsymbol{w}^{\mathsf{MF}*}_k\boldsymbol{h}_k\right|^2\right] = \alpha_u^2 G_k P_k(N_{\mathsf{a}}^2 + N_{\mathsf{a}})
\end{equation}
while the first term of (\ref{den1}) vanishes since the CSI is perfect.
\begin{proposition}
Considering a single-cell multiuser system (without any inter-cell interference) with perfect CSI, Corollary \ref{cor2} entails the results for uplink users in \cite{massiveadc}.
\end{proposition}

\section{Forward Link (Downlink) Analysis}
Before the DAC, the signal transmitted by the $\ell$-th BS is
\begin{equation}
\boldsymbol{x}_{\ell} = \sum_{k=0}^{K_\ell^d-1} \sqrt{\frac{P_{\ell,k}}{N_{\mathsf{a}}}}\boldsymbol{f}_{\ell,k}s_{\ell,k}
\end{equation}
where $P_{\ell,k}$ is the power allocated to the data symbol $s_{\ell,k} \sim \mathcal{N}_{\mathbb{C}}(0,1)$, which is precoded by $\boldsymbol{f}_{\ell,k}$ and intended for its $k$-th user. The power allocation satisfies 
\begin{equation}
    \sum_{k=0}^{K_\ell^d-1}P_{\ell,k} = P
\end{equation}
where $P$ is the downlink power per cell. Since the BS operates in FD mode, uplink users are corrupted by SI while the downlink users are not. Since uplink users are transmitting at the same time downlink users are receiving, the latter are vulnerable to inter-user interference from uplink users.

%% Upon data transmission from the BS of interest,
The $k$-th user observes the following quantized received signal from the BS of interest as
\begin{equation}
\begin{split}\label{downlink1}
y_{q,k}^d =&   \alpha_d \sum_{\ell}\sum_{\text{k}=0}^{K_{\ell}^d-1} \sqrt{\frac{G_{\ell,k}P_{\ell,\text{k}}}{N_{\mathsf{a}}}} \boldsymbol{h}^*_{\ell,k}\boldsymbol{f}_{\ell,\text{k}}s_{\ell,\text{k}} \\&+ \sum_{\ell}\sum_{\text{k}=0}^{K_\ell^d-1} \sqrt{\frac{G_{\ell,k}P_{\ell,\text{k}}}{N_{\mathsf{a}}}}\boldsymbol{h}^*_{\ell,k} \boldsymbol{q}_{d,\ell}\\&+ \sum_{\ell}\sum_{\text{k}=0}^{K_{\ell}^u-1} \sqrt{T_{(\ell,\text{k}),k}P_{\ell,\text{k}}^u} \boldsymbol{g}_{(\ell,\text{k}),k} s_{\ell,\text{k}}^u  +v_k
\end{split}    
\end{equation}

\subsection{Channel Hardening}
Since we consider receivers reliant on channel hardening, the $k$-th user served by the BS of interest regards $\mathbb{E}[\boldsymbol{h}^*_{k}\boldsymbol{f}_k]$ as its precoded channel wherein the small-scale fading is averaged. The variation of the actual precoded channel around the mean incurs SI, such that (\ref{downlink1}) can be formulated as (\ref{downlink2}).
\begin{figure*}[t]
\begin{equation}\label{downlink2}
\begin{split}
    &y_{q,k}^d = \underbrace{\alpha_d \sqrt{\frac{G_{k}P_k}{N_{\mathsf{a}}}} \mathbb{E}[\boldsymbol{h}^*_{(k)}\boldsymbol{f}_k]s_k}_{\textsf{Desired Signal}} + \underbrace{\alpha_d \sqrt{\frac{G_{k}P_k}{N_{\mathsf{a}}}}\left( \boldsymbol{h}^*_{k}\boldsymbol{ f}_k -  \mathbb{E}[\boldsymbol{h}^*_{k}\boldsymbol{f}_k] \right) s_k}_{\textsf{Channel Estimation Error (Self-Interference)}} + 
    \underbrace{\alpha_d \sum_{\text{k}\neq k} \sqrt{\frac{G_{k}P_{\text{k}}}{N_{\mathsf{a}}}} \boldsymbol{h}^*_{k}\boldsymbol{f}_{\text{k}}s_{\text{k}}}_{\textsf{Intra-Cell Interference}} +     \underbrace{ \sum_{\ell}\sum_{\text{k}=0}^{K^d_\ell-1} \sqrt{\frac{G_{\ell,k}P_{\ell,\text{k}}}{N_{\mathsf{a}}}}\boldsymbol{h}^*_{\ell,k} \boldsymbol{q}_{\ell}}_{\textsf{Aggregate AQNM}} \\&+ \underbrace{\alpha_d \sum_{\ell\neq 0}\sum_{\text{k}=0}^{K_{\ell}^d-1} \sqrt{\frac{G_{\ell,k}P_{\ell,\text{k}}}{N_{\mathsf{a}}}} \boldsymbol{h}^*_{\ell,k}\boldsymbol{f}_{\ell,\text{k}}s_{\ell,\text{k}}}_{\textsf{Inter-Cell Interference}} + \underbrace{\sum\limits_{\text{k}\neq k} \sqrt{T_{\text{k},k}P_{\ell,\text{k}}^u} \boldsymbol{g}_{\text{k},k} s_{\text{k},u}}_{\textsf{Same Cell Inter-User Interference}}+ \underbrace{\sum_{\ell \neq 0}\sum\limits_{\text{k}=0}^{K_{\ell}^u-1} \sqrt{T_{(\ell,\text{k}),k}P_{\ell,\text{k}}^u} \boldsymbol{g}_{(\ell,\text{k}),k} s_{\ell,\text{k},u}}_{\textsf{Other Cells Inter-User Interference}}  
 +  \underbrace{v_k}_{\textsf{Noise}}
    \end{split}    
\end{equation}
\vspace*{-.5cm}
\end{figure*}

\subsection{Matched Filter Precoder}
From the reverse link pilots transmitted by its users,
the $\ell$-th BS gathers channel estimates $\hat{\boldsymbol{h}}_{\ell,(\ell,0)},\ldots,\hat{\boldsymbol{h}}_{\ell,(\ell,K_\ell^d-1)}$. With matched filter transmitter, the precoders at cell $\ell$ are given by
\begin{equation}
\boldsymbol{f}_{\ell,k}^{\mathsf{MF}} = \sqrt{N_{\mathsf{a}}}\frac{\hat{\boldsymbol{h}}_{\ell,(\ell,k)}}{\sqrt{\mathbb{E}\left[ \left\|  \hat{\boldsymbol{h}}_{\ell,(\ell,k)} \right\|^2 \right]}},~k=0,\ldots,K_\ell^d-1    
\end{equation}
where the precoders share the same properties as the matched filter receiver indicated by Corollary \ref{cor1}.
\begin{theorem}
The output SQINR of the $k$-th downlink user is 
\begin{equation}\label{downlinksqinr}
\overline{\mathsf{sqinr}}_k^{\mathsf{MF}} = \frac{ \alpha_d^2 \frac{G_k P_k}{N_{\mathsf{a}}}  \left|\mathbb{E}\left[\boldsymbol{h}^*_k\boldsymbol{f}^{\mathsf{MF}}_k\right]\right|^2  }{\overline{\mathsf{den}}^{\mathsf{MF}}_d}    
\end{equation}
\end{theorem}
where $\overline{\mathsf{den}}^{\mathsf{MF}}_d$ is given by (\ref{den2}).
\begin{figure*}[t]
\begin{equation}\label{den2}
\begin{split}
\overline{\mathsf{den}}^{\mathsf{MF}}_d =&  \alpha_d^2 \frac{G_kP_k}{N_{\mathsf{a}}}\textsf{var}\left[ \boldsymbol{h}^*_k\boldsymbol{f}^{\mathsf{MF}}_k \right] +\alpha_d^2 \sum_{\text{k}\neq k}\frac{G_kP_{\text{k}}}{N_{\mathsf{a}}} \mathbb{E}\left[\left|\boldsymbol{h}_k^* \boldsymbol{f}_{\text{k}}^{\mathsf{MF}}  \right|^2 \right] +\alpha_d^2 \sum_{\ell \neq 0}\sum_{\text{k}=0}^{K_\ell^d-1}\frac{G_{\ell,k}P_{\ell,\text{k}}}{N_{\mathsf{a}}}\mathbb{E}\left[ \left| \boldsymbol{h}_{\ell,k}^* \boldsymbol{f}_{\ell,\text{k}}  \right|^2  \right]  \\&+ \sum_{\text{k}\neq k}T_{\text{k},k}P_{\text{k}}^u\mathbb{E}\left[ \left| \boldsymbol{g}_{\text{k},k} \right|^2 \right]+ \sum_{\ell \neq 0}\sum_{\text{k}=0}^{K_\ell^u-1}T_{(\ell,\text{k}),k}P_{\ell,\text{k}}^u\mathbb{E}\left[ \left| \boldsymbol{g}_{(\ell,\text{k}),k} \right|^2 \right] + \sum_{\ell} \frac{G_{\ell,k}P_{\ell,k}}{N_{\mathsf{a}}} \mathbb{E}\left[ \left| \boldsymbol{h}^*_{\ell,k}\boldsymbol{q}_{\ell} \right|^2 \right] + \sigma^2
\end{split}
\end{equation}
\vspace*{-.5cm}
\end{figure*}

The numerator in (\ref{downlinksqinr}) and the first three terms of (\ref{den2}) can be solved using Corollary \ref{cor1}, while the remaining terms in (\ref{den2}) can be derived as
\begin{equation}\label{b1}
\sum_{\ell}\sum_{\text{k}=1}^{K_{\ell}^u} T_{(\ell,\text{k}),k}P_{\ell,\text{k}}^u\mathbb{E}\left[ \left| \boldsymbol{g}_{(\ell,\text{k}),k} \right|^2 \right] = \sum_{\ell}\sum_{\text{k}=1}^{K_{\ell}^u} T_{(\ell,\text{k}),k}P^u_{\ell,\text{k}} \sigma^2_{\mathsf{iui}}
\end{equation}
\begin{equation}\label{b2}
\sum_{\ell} \frac{G_{\ell,k}P_{\ell,k}}{N_{\mathsf{a}}} \mathbb{E}\left[ \left| \boldsymbol{h}^*_{\ell,k}\boldsymbol{q}_{\ell}^d \right|^2 \right] =  \alpha_d(1-\alpha_d)\sum_{\ell}G_{\ell,k}P_{\ell,k} (K_\ell^d+1)  
\end{equation}

\begin{proof}
The proof of the (\ref{b1}) and (\ref{b2}) are given by \cite{massiveadc}.
\end{proof}
Assuming perfect CSI and applying Corollary \ref{cor2}, the numerator of (\ref{downlinksqinr}) becomes
\begin{equation}
\overline{\text{num}}^{\mathsf{MF}}_d = \alpha_d^2\frac{G_kP_k}{N_{\mathsf{a}}}\mathbb{E}\left[ \left|  \boldsymbol{h}^*_k \boldsymbol{f}_k^{\mathsf{MF}} \right|^2  \right] = \alpha_d^2G_kP_k(N_{\mathsf{a}}+1) 
\end{equation}
while the first term of (\ref{den2}) vanishes since the CSI is perfect.
\begin{proposition}
Considering a single-cell multiuser system (without any inter-cell interference) with perfect CSI, Corollary \ref{cor2} entails the results for downlink users in \cite{massiveadc}.
\end{proposition}
\begin{proposition}
For channel hardening without full-duplexing (hence no co-channel interference between users) and with full-resolution ADC/DACs, we retrieve the results derived for downlink users in multicell massive MIMO systems in \cite{massiveforward}. 
\end{proposition}

\section{Asymptotic Analysis and Power Scaling Laws}
In the section, we investigate the effects of the number of quantization bits, the number of antennas at the BS, the power budgets of the BS and each user on the spectral efficiency performance for reverse and forward links.
%\subsection{Reverse Link Analysis}
\begin{lemma}
For a fixed power budget, fixed number of transmit antennas and full-resolution ($b \rightarrow \infty$, $\alpha_u = \alpha_d = 1$), the spectral efficiencies for reverse and forward links converge to
\begin{equation}\label{uplinklemma1}
\frac{\bar{\mathcal{I}}_k^u}{B} \rightarrow \log\left(1+  \frac{G_kP_kN_{\mathsf{a}}}{ \sum_{\ell}\sum_{\text{k}}G_{\ell,\text{k}}P_{\ell,\text{k}}+ P_{\mathsf{SI}}\mu^2_{\mathsf{SI}}K^dN_{\mathsf{a}} + \sigma^2} \right)   
\end{equation}
\begin{equation}\label{downlinklemma1}
\begin{split}
&\frac{\bar{\mathcal{I}}_k^d}{B} \rightarrow \\&\log\left(1+  \frac{G_kP_k}{\sum_\ell\sum_{\text{k}}G_{\ell,\text{k}}P_{\ell,\text{k}} + \sum_{\ell}\sum_{\text{k}}T_{(\ell,\text{k}),k}P_{\ell,\text{k}}^u \sigma_{\mathsf{iui}}^2 + \sigma^2 }\right)   
\end{split}
\end{equation}
\end{lemma}

\noindent
In (\ref{uplinklemma1}) and (\ref{downlinklemma1}), the quantization error incurred by ADC/DAC is removed for full-resolution. When the number of antennas is fixed, the spectral efficiency for up/downlink users becomes constant. Although increasing ADC/DAC resolution can enhance the performance, the rate is still limited.
\begin{lemma}
For a fixed number of antennas, fixed $b$ and when $P_{\mathsf{SI}} = P^d = P^u \rightarrow \infty$, the spectral efficiency converges to 
\begin{equation}
\begin{split}\label{uplinklemma2}
\frac{\bar{\mathcal{I}}_k^u}{B} \rightarrow \log\left( 1 + \frac{\alpha_uG_kN_{\mathsf{a}}}{ \splitfrac{  \sum_\ell\sum_{\text{k}} G_{\ell,\text{k}}+(1-\alpha_u)G_k}{ +\alpha_dN_{\mathsf{a}}\mu_{\mathsf{SI}}^2  \left[ 1+\alpha_u\alpha_d(K^d-1) \right] }}   \right)  
\end{split}
\end{equation}
\begin{equation}\label{downlinklemma2}
\begin{split}
\frac{\bar{\mathcal{I}}_k^d}{B} \rightarrow \log\left(1+  \frac{\alpha_d^2G_k}{ \splitfrac{ \alpha_d^2\sum_\ell\sum_{\text{k}}G_{\ell,\text{k}} + \sum_{\ell}\sum_{\text{k}}T_{(\ell,\text{k}),k} \sigma_{\mathsf{iui}}^2} {+\alpha_d(1-\alpha_d)\sum_\ell G_{\ell,k}(K_\ell^d+1)}}\right)   
\end{split}
\end{equation}
\end{lemma}
\noindent
In (\ref{uplinklemma2}), the uplink spectral efficiency depends on number of antennas $N_{\mathsf{a}}$ and quantization bits $b$ when BS transmit power and number of users go to infinity. Spectral efficiency saturates by a ceiling caused by SI and quantization error.
\begin{lemma}
If the transmit powers of the BS and each user is scaled with the number of antennas $N_{\mathsf{a}}$ i.e., $P = \frac{E}{N_{\mathsf{a}}}$ where $E$ is fixed, as $N_{\mathsf{a}} \rightarrow \infty$, the spectral efficiency converges to 
\begin{equation}\label{uplinklemma3}
\frac{\bar{\mathcal{I}}_k^u}{B} \rightarrow \log\left(1+ \frac{
\alpha_uG_kE_k}{\alpha_d\mu_{\mathsf{SI}}^2E_{\mathsf{SI}} \left[1+\alpha_u\alpha_d(K^d-1)\right]+  \sigma^2}\right)
\end{equation}
\end{lemma}
We found that using a proper power scaling law and more antennas can eliminate the intra-cell and inter-cell interference. The number of quantization bits provides an approximation of the uplink spectral efficiency when $N_{\mathsf{a}}$ goes to infinity. 
\begin{equation}\label{downlinklemma3}
\frac{\bar{\mathcal{I}}_k^d}{B} \rightarrow \log\left(1+ \frac{
\alpha^2_dG_kE_k}{\sigma^2}\right) 
\end{equation}
(\ref{downlinklemma3}) implies that using a proper power scaling and more antennas can eliminate the inter-user interference caused by full-duplexing. The number of quantization bits determines the approximate downlink spectral efficiency when the number of the antennas at a FD BS, $N_{\mathsf{a}}$, goes to infinity.

\section{Numerical Results}
%% In this section, we provide the numerical results of the system performance along with the discussion. 
Simulations use the values below unless otherwise stated. We assume uniform power allocation for forward links.

\vspace*{0.1in}
%\begin{table}[b]
%\renewcommand{\arraystretch}{1}
%\caption{System Parameters %\cite{values,massiveforward,massiveadc}.}
%\label{sysparam}
%\centering
\begin{tabular}{rl}
%\hline
%\textbf{Parameter} & \textbf{Value}\\
%\hline
Bandwidth & 20 MHz\\
%BS Density ($\lambda_{\text{BS}}$) & 0.1 BS/$\text{km}^2$\\
%UE Density ($\lambda_{\text{UE}}$) & 0.5 UE/$\text{km}^2$\\
%Pathloss Intercept ($L_{\text{ref}}$) & 37$\log(r)$, $r$: distance in meter\\
Pathloss Exponent & 3.5\\
Shadowing & 5 dB \\
Downlink Transmit Power & 40 W\\
Uplink Transmit Power & 250 mW\\
SI Power & 40 W\\
SI Channel Power & 10 dB \\
Noise Spectral Density & -174 dBm/Hz\\
Number of antennas & 100
%\hline
\end{tabular}
%\end{table}
\vspace*{0.1in}

\begin{figure}[t]
\centering
\setlength\fheight{5.5cm}
\setlength\fwidth{7.3cm}
\input{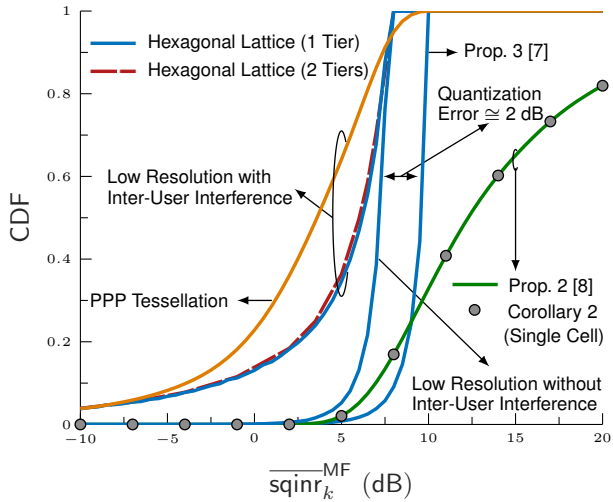}
    \caption{Forward link results: Effects of full-duplexing, quantization error, and network cell shapes on outage probability (CDF) with $\alpha_u = \alpha_d = 0.6$.}
    \label{pict1}
\end{figure}

Fig.~\ref{pict1} plots the cumulative distribution function (CDF) or outage probability of the downlink SQINR.
In the CDF, the PPP tessellation exceeds the hexagonal lattice except for a small range of 8--12 dB of downlink SNR.
%% The performance for PPP tessellation serves as an upper bound for the hexagonal lattice. This justifies that PPP is more accurate because it takes into account the network degradation not captured by using the hexagonal grid. 
In addition, the performance decreases with increasing inter-user interference incurred by FD operation of the BS and uplink UEs transmission. The plot shows the loss incurred by the low-resolution ADC/DAC is about 2 dB when the CDF saturates.

\begin{figure}[t]
\centering
\setlength\fheight{5.5cm}
\setlength\fwidth{7.3cm}
\input{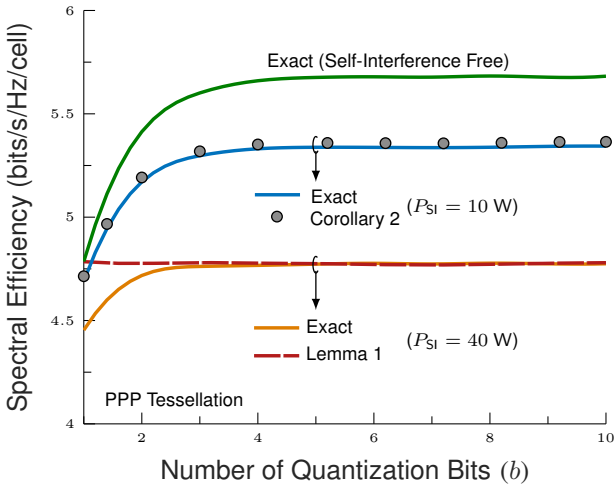}
    \caption{Reverse link spectral efficiency vs. number of quantization bits and SI power.}
    \label{pict2}
\end{figure}

Fig. \ref{pict2} simulates uplink spectral efficiency vs. number of quantization bits ($b$) with a PPP tessellation network. Spectral efficiency increases with $b$ and converges to a ceiling derived by Lemma 1 (\ref{uplinklemma1}); the rate decreases when adopting low-resolution ADC/DACs (low $b$). Spectral efficiency degrades for 10W and 40W of SI power vs. the SI-free case.
\begin{figure}[t]
\centering
\setlength\fheight{5.5cm}
\setlength\fwidth{7.3cm}
\input{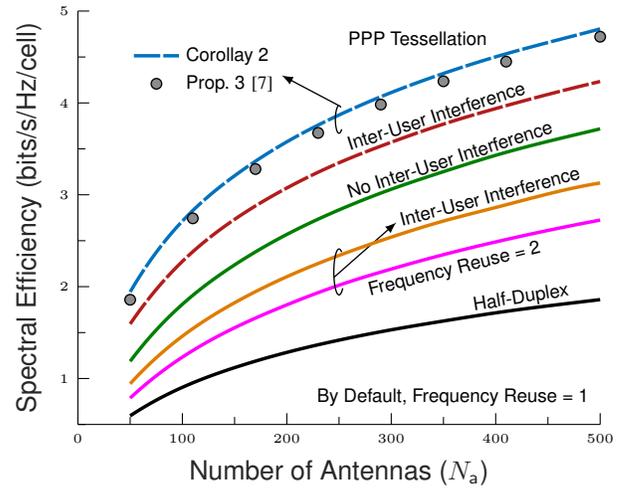}
    \caption{Forward link results: Effects of the number of antennas, duplexing modes, ADC/DAC resolution and frequency reuse factor on spectral efficiency with $\alpha_u = \alpha_d = 0.5$. \textcolor{black}{Solid and dashed lines stand for low and full resolution ADC/DACs, respectively}.}
    \label{pict4}
\end{figure}

Fig. \ref{pict4} shows downlink spectral efficiency
increasing with the number of quantization bits and BS antennas for a PPP tessellation. When considering low- and full-resolution  data converters, spectral efficiency converges to a fixed ceiling with enough antennas, which agrees with Lemma 3 (\ref{downlinklemma3}). 
%% Given that, increasing frequency reuse is often adopted to improve cellular coverage with a tradeoff of decreasing the rate in the network. 

\section{Conclusion}
In this work, we show the feasibility of FD in massive MIMO cellular networks with low-resolution ADCs and DACs. Using matched filter precoding and combining at the BS, AQNM modeling for the DACs and ADCs, and channel hardening, we analyzed the SQINR CDF and spectral efficiency for reverse and forward links for hexagonal and PPP networks. Using proper power scaling and enough antennas can eliminate intra-cell, inter-cell and FD inter-user interference. Simulation results indicate quantization error and SI cause pronounced losses in spectral efficiency; however, this loss can be compensated by using more antennas.
%\balance
\bibliographystyle{IEEEtran}
\bibliography{main}

\begin{thebibliography}{10}
\providecommand{\url}[1]{#1}
\csname url@samestyle\endcsname
\providecommand{\newblock}{\relax}
\providecommand{\bibinfo}[2]{#2}
\providecommand{\BIBentrySTDinterwordspacing}{\spaceskip=0pt\relax}
\providecommand{\BIBentryALTinterwordstretchfactor}{4}
\providecommand{\BIBentryALTinterwordspacing}{\spaceskip=\fontdimen2\font plus
\BIBentryALTinterwordstretchfactor\fontdimen3\font minus
  \fontdimen4\font\relax}
\providecommand{\BIBforeignlanguage}[2]{{%
\expandafter\ifx\csname l@#1\endcsname\relax
\typeout{** WARNING: IEEEtran.bst: No hyphenation pattern has been}%
\typeout{** loaded for the language `#1'. Using the pattern for}%
\typeout{** the default language instead.}%
\else
\language=\csname l@#1\endcsname
\fi
#2}}
\providecommand{\BIBdecl}{\relax}
\BIBdecl

\bibitem{release17}
``{The 5G Evolution: 3GPP Rel. 16-17},''
  \emph{https://www.5gamericas.org/wp-content/uploads/2020/01/5G-Evolution-3GPP-R16-R17-FINAL.pdf}.

\bibitem{ianmagazine}
I.~P. Roberts, J.~G. Andrews, H.~B. Jain, and S.~Vishwanath, ``Millimeter-wave
  full duplex radios: New challenges and techniques,'' \emph{IEEE Wireless
  Communications}, vol.~28, no.~1, pp. 36--43, 2021.

\bibitem{zf}
E.~Balti and N.~Mensi, ``Zero-forcing max-power beamforming for hybrid {mmWave}
  full-duplex {MIMO} systems,'' in \emph{Int. Conf. Adv. Systems and Emergent
  Technologies}, 2020, pp. 344--349.

\bibitem{f2}
A.~Koc and T.~Le-Ngoc, ``Full-duplex mmwave massive mimo systems: A joint
  hybrid precoding/combining and self-interference cancellation design,''
  \emph{IEEE Open Journal of the Communications Society}, vol.~2, pp. 754--774,
  2021.

\bibitem{ianjournal}
I.~P. Roberts, J.~G. Andrews, and S.~Vishwanath, ``Hybrid beamforming for
  millimeter wave full-duplex under limited receive dynamic range,'' \emph{IEEE
  Trans. Wireless Commun.}, vol.~20, no.~12, pp. 7758--7772, 2021.

\bibitem{overview}
R.~W. {Heath}, N.~{González-Prelcic}, S.~{Rangan}, W.~{Roh}, and A.~M.
  {Sayeed}, ``An overview of signal processing techniques for millimeter wave
  {MIMO} systems,'' \emph{IEEE J. Selected Topics in Signal Process.}, vol.~10,
  no.~3, pp. 436--453, 2016.

\bibitem{massiveforward}
G.~George, A.~Lozano, and M.~Haenggi, ``Massive {MIMO} forward link analysis
  for cellular networks,'' \emph{IEEE Trans. Wireless Commun.}, vol.~18, no.~6,
  p. 2964–2976, Jun 2019.

\bibitem{massiveadc}
J.~Dai, J.~Liu, J.~Wang, J.~Zhao, C.~Cheng, and J.-Y. Wang, ``Achievable rates
  for full-duplex massive {MIMO} systems with low-resolution {ADCs/DACs},''
  \emph{IEEE Access}, vol.~7, pp. 24\,343--24\,353, 2019.

\bibitem{R3}
Q.~Ding, Y.~Lian, and Y.~Jing, ``Performance analysis of full-duplex massive
  mimo systems with low-resolution {ADCs/DACs} over {Rician} fading channels,''
  \emph{IEEE Trans. Veh. Tech.}, vol.~69, no.~7, 2020.

\bibitem{fdvehicular}
P.~Anokye, R.~K. Ahiadormey, and K.-J. Lee, ``Full-duplex cell-free massive
  mimo with low-resolution adcs,'' \emph{IEEE Transactions on Vehicular
  Technology}, vol.~70, no.~11, pp. 12\,179--12\,184, 2021.

\bibitem{R5}
X.~Yu, J.~Dai, and J.~Shi, ``Achievable rates of full-duplex massive {MIMO}
  systems with {Mixed-ADC/DAC},'' in \emph{IEEE Int. Conf. High Perf. Computing
  and Communications}, 2019, pp. 1692--1698.

\bibitem{R6}
P.~Anokye, R.~K. Ahiadormey, H.-S. Jo, C.~Song, and K.-J. Lee, ``Low-resolution
  {ADC} quantized full-duplex massive {MIMO}-enabled wireless backhaul in
  heterogeneous networks over {Rician} channels,'' \emph{IEEE Trans. Wireless
  Commun.}, vol.~19, no.~8, pp. 5503--5517, 2020.

\bibitem{R7}
P.~Anokye, R.~K. Ahiadormey, C.~Song, and K.-J. Lee, ``On the sum-rate of
  heterogeneous networks with low-resolution {ADC} quantized full-duplex
  massive {MIMO}-enabled backhaul,'' \emph{IEEE Wireless Commun. Let.}, vol.~8,
  no.~2, pp. 452--455, 2019.

\bibitem{lowres}
X.~Zhang, T.~Liang, K.~An, G.~Zheng, and S.~Chatzinotas, ``Secure transmission
  in cell-free massive mimo with rf impairments and low-resolution adcs/dacs,''
  \emph{IEEE Transactions on Vehicular Technology}, vol.~70, no.~9, pp.
  8937--8949, 2021.

\end{thebibliography}
\end{document}